\documentclass[lettersize,journal]{IEEEtran}

\usepackage{array}
\usepackage[caption=false,font=normalsize,labelfont=sf,textfont=sf]{subfig}
\usepackage{textcomp}
\usepackage{stfloats}
\usepackage{url}
\usepackage{verbatim}
\usepackage{graphicx}
\usepackage{cite}
\usepackage{amsmath,amssymb,amsfonts}
\usepackage{algorithmic}
\usepackage{algorithm}
\usepackage{graphicx}
\usepackage{float}
\usepackage{footmisc}
\usepackage{stfloats}
\usepackage{textcomp}
\usepackage{multirow}
\usepackage{lscape}
\usepackage{caption}
\usepackage{enumerate}
\usepackage{xcolor}
\usepackage{flushend}
\begin{document}

\title{Liquid Crystal-Based RIS for VLC Transmitters: Performance Analysis, Challenges, and Opportunities}

\author{Sylvester Aboagye, Telex~M.~N.~Ngatched, Alain R. Ndjiongue, Octavia~A.~Dobre,~and Hyundong Shin 
\thanks{This paper has been accepted for publication by IEEE. Copyright may be transferred without notice, after which this version may no longer be accessible.}
}



\maketitle

\begin{abstract}
This article presents a novel approach of using reconfigurable intelligent surfaces (RISs) in the transmitter of indoor visible light communication (VLC) systems to enhance data rate uniformity and maintain adequate illumination. In this approach, a liquid crystal (LC)-based RIS is placed in front of the LED arrays of the transmitter to form an LC-based RIS-enabled VLC transmitter. This RIS-enabled transmitter is able to perform new functions such as transmit light steering and amplification and demonstrates very high data rate and illumination performance when compared with traditional VLC transmitters with circular and distributed LED arrays and the more recent angle diversity transmitter. Simulation results reveal the strong potential of LC-based RIS-aided transmitters in satisfying the joint illumination and communication needs of indoor VLC systems and positions VLC as a critical essential block for next generation communication networks. Several challenging and exciting issues related to the realization of such transmitters are discussed.
\end{abstract}


\section{Introduction}
\IEEEPARstart{V}{isible} light communication (VLC) has emerged as one of the key revolutionary technologies to support energy-efficient, secure, and high data rate transmissions with low deployment costs in the next generation of communication networks \cite{9968053}. This is because of the huge and unlicensed bandwidth availability in the visible light spectrum and the rapid development of light-emitting diodes (LEDs), which serve as transmitters in VLC. However, the practical deployment of VLC systems, especially in indoor environments, has been faced with unique challenges such as non-uniform illumination and data rate coverage \cite{7096280,Wang12,9799770,9037594}, line-of-sight (LoS) blockages, 
 random device orientation, and loss of incident signal power in VLC receivers \cite{9354893}. The focus of this article relates to the non-uniform illumination and data rate coverage design issue. Although this challenge can be considered the most important since there can be no successful communication without adequate network coverage, it has received the least attention. Specifically, large variations in illumination and data rate, especially in the corners of indoors, limit the system's ability to provide ubiquitous services and high data rates to multiple users located at different places. Moreover, the quality of any user's experience should not be defined solely or affected by the location in a room as all users should enjoy high communication quality.

The authors in \cite{7096280} examined the illumination properties of different LED array arrangements and experimentally demonstrated the communication and illumination performance of a phosphor-based VLC system. In \cite{Wang12}, the authors investigated the bit error rate and channel capacity performance of circular and centered LED arrangements. In \cite{9037594}, the authors analyzed the effects of LED array layout on the illumination uniformity of VLC systems. An optimization problem to determine the optimal placement of LED arrays to maximize the average area spectral efficiency has been considered in \cite{6691890}. The results indicated that LED arrays must be deployed in the middle of the room to support high spectral efficiency and illumination. The authors in \cite{9578932,7857700,9614037}  studied the use of angle diversity transmitters (ADTs) to reduce illumination fluctuation and provide a more uniform data rate in indoor VLC systems.   

Recently, a number of studies such as \cite{9968053,9799770,9526581,9500409,9910023,9354893} have investigated the use of optical reconfigurable intelligent surfaces (RISs) to solve the design problems (i) to (iv). In \cite{9799770}, the authors proposed a novel approach of using mirrors to enhance the illumination uniformity and data rate of an indoor multi-cell VLC system. The studies in \cite{9968053,9526581,9500409} proposed RIS-aided VLC system models to combat the blockage and random device orientation issues in an indoor environment. In \cite{9354893,9910023}, the authors examined a liquid crystal (LC)-based RIS to enhance signal detection in VLC receivers by performing incident light amplification and light steering. In the studies mentioned above, the considered system models involved the deployment of optical RISs either in the transmission channel \cite{9968053,9526581,9500409,9799770} or in front of the photodetector (PD) of the VLC receiver \cite{9354893,9910023}. To the best of the authors' knowledge, there has yet to be a study on the application of optical RISs at the transmitter side to boost the coverage and enhance key performance metrics of VLC systems.

In VLC systems, it is highly desirable to ensure illumination uniformity and high speed data transmission in the indoor environment. This is necessary to ensure that quality-of-service is not dependent on the location of the user. However, the current approaches of deploying light fixtures (e.g., LEDs or LED arrays) indoors do not achieve uniform illumination coverage as, most often, the corners of various rooms have less illumination. An LED with a large semi-angle can generate wider beam angles to ensure more uniform illumination indoors. However, the generation of such broad beams results in a decrease in the intensity of the optical signals. Only a few studies have investigated methods of providing even illumination and high data rates in indoor VLC systems. Furthermore, the impact of LED/LED array arrangement on the performance of indoor VLC systems has yet to receive much attention. Note that the LED array arrangement can play a vital role in the overall system performance of VLC. Unlike \cite{9354893} and \cite{9910023}, for the first time this article explores the use of LC-based RISs in the transmitter of a VLC system to jointly improve the communication and illumination performances in indoor VLC systems. In this article, a novel approach of using LC-based RIS to enhance illumination  and data rate uniformity of an indoor VLC system is presented, bringing out the following main contributions:  

\begin{itemize}
    \item We propose a novel application of optical RISs in VLC systems. Specifically, an LC-based RIS-enabled transmitter design is introduced and the impact of the RIS on the emerging light from the LED is analyzed.
    \item We examine the propagation characteristics of a VLC system with RIS-enabled transmitters as optical signals travel from the LEDs to the receivers using geometric optics. In addition, we provide an expression to characterize the illumination distribution.
    \item We use simulations to quantify the potential gains of deploying RISs in front of LED arrays and present a performance comparison with a VLC system equipped with ADTs and other popular LED array arrangement schemes.
    \item Finally, we discuss other potential applications of this novel transmitter design and present several exciting and challenging research opportunities that can further improve its performance and accelerate its realization for future generation optical wireless networks.
\end{itemize}

\section{Traditional Deployment of LEDs Indoors}
This section describes the various and modern ways of LED array placement indoors. Although the deployment of indoor light fixtures could depend on design factors such as size and space, occupant's age and preference, ceiling height and shape, to mention only a few, a general description is provided. 

\begin{figure}
    \centering
    {{\includegraphics[width=0.4\textwidth]{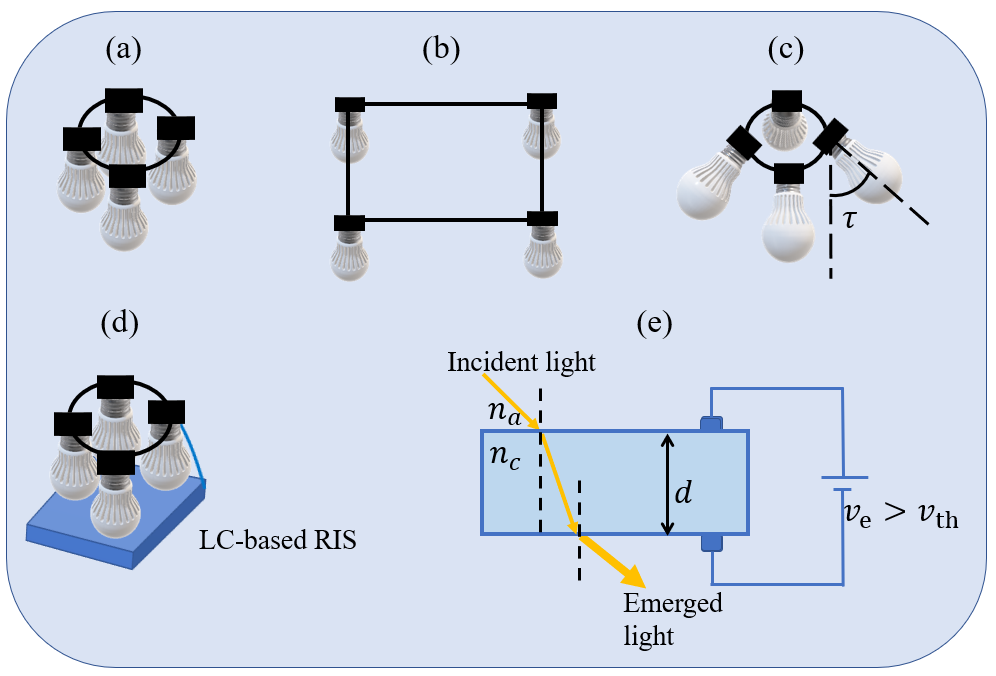}}}
    \caption{\small{Typical LED arrangement indoors and the proposed LC-based RIS-enabled VLC transmitter: (a) centralized LED array placement; (b) distributed LED array placement; (c) ADT LED array placement; (d) centralized LED arrays with an LC-based RIS at the front end; (e) geometry of light signal propagation and amplification through the LC-based RIS element.}}%
  \label{figs01}
   \vspace*{-3mm}
\end{figure}

As the primary purpose of LED arrays is to provide sufficient indoor illumination, their traditional deployment has always focused solely on lighting and illumination. Due to the Lambertian radiation pattern of LEDs and the fact that most human interactions or conversations occur at the center of indoor environments (i.e., the task or activity area), LED arrays are typically deployed as ceiling fixtures at/or near the center to provide adequate functional illumination. Such a typical LED array arrangement is illustrated in Fig.~\ref{figs01} (a) where, under the point source assumption, 4 LED arrays are deployed as a ceiling light fixture that directs light downwards. Figure~\ref{figs01} (b) depicts another popular LED array arrangement where the 4 LED arrays are distributed on the ceiling in a rectangular/square shape, with each LED array serving as a point source transmitter. The LED array arrangements in Figs.~\ref{figs01} (a) and (b) have been extensively considered in earlier studies on indoor VLC systems such as  \cite{Wang12,7096280,9614037,9799770,9500409,9910023,6691890,9526581}.



A more recent approach for LED array placement, referred to as ADT arrays, is shown in Fig.~\ref{figs01} (c). In this figure, each LED array is inclined at an elevation angle $\tau$ to point toward a particular direction to improve the illumination level throughout. ADTs have been considered in the design of multi-cell VLC systems in a few recent studies (e.g., \cite{9578932,7857700,9614037}). These studies demonstrated the ADTs superior data rate and energy efficiency performance compared to the centralized and distributed LED array placement schemes. However, illumination and data rate uniformity assessments have yet to be reported on ADTs. 

\section{LC-Based RIS-Enabled VLC System}
\subsection{LC-Based RIS-Enabled Transmitter Design}
Figure~\ref{figs01} (d) depicts the proposed RIS-enabled VLC transmitter design. In this figure, an LC-based RIS is deployed in front of a centralized LED array arrangement.\footnote{Note that the discussions on channel gain expression and analysis on the performance for this transmitter design applies to other LED array arrangements such as ADT and distributed LED arrays.} The motivation for considering LC-based RIS is its light steering and amplification capabilities when subjected to an external electric field. More specifically, LC-based RISs are characterized by their electronically tunable physico-chemical properties (e.g., refractive index, emission, and attenuation coefficients) that can be easily controlled by the arrangement of the LC molecules via an external electric field. Moreover, LCs, in general, have received significant interest in optical wireless networks and have been considered in the development of optical filters for communications, next generation light detection and ranging sensors for self-driving vehicles, and optical receivers \cite{9354893}. 

Figure~\ref{figs01} (e) illustrates the principles of light steering and amplification in the LC-based RIS when an external voltage, $v_{\rm e}$, greater than the threshold voltage, $v_{\rm th}$, is applied. The threshold voltage is a critical voltage at which the re-orientation of the LC's molecules begins. As shown in this figure, the emitted light beam from the LED arrays impinges on the LC-based RIS. At the surface of the LC-based RIS, part of the incident light beam gets reflected while the remaining beam undergoes refraction as it passes from the air medium with a refractive index $n_a$ into the LC element with a refractive index $n_c$. Inside the LC-based RIS element, the light beam's photons interact with the LC's excited molecules (due to the presence of the external electric field). This causes the excited molecules to drop to a lower energy level, resulting in the generation of identical new photons through the principle of stimulated emission. The direction of the resulting light beam (i.e., incoming photons and the generated photons) as it propagates through and exits the LC-based RIS element with thickness $d$ can be controlled through an electric field-induced molecular reorientation \cite{9910023,9354893}. More particularly, the transmission coefficient that characterizes light propagation in the LC-based RIS element is a function of the refractive index $n_c$. This refractive index can be tuned electronically by changing the orientation of the LC-based RIS element's molecules through the external voltage $v_{\rm e}$. Thus, incident light steering and amplification are obtained by subjecting the LC-based RIS to the external electric field.    

\begin{figure}
    \centering
    {{\includegraphics[width=0.4\textwidth]{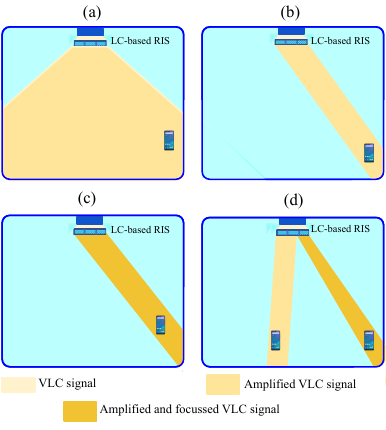}}}
    \caption{\small{Application scenarios of the proposed LC-based RIS-enabled VLC transmitter: (a) transmit signal amplification and coverage enhancement; (b) dynamic beam steering; (c) transmit signal amplification and beam focusing; (d) secured and interference-free VLC.}}%
  \label{adt}
  \vspace*{-4mm}
\end{figure}

Figure~\ref{adt} illustrates the various application scenarios of the LC-based RIS-enabled VLC transmitter for indoor VLC systems. Figure~\ref{adt} (a) shows how the LC-based RIS can be used to amplify incoming light from  LEDs to produce high-intensity emitted visible light signal to jointly enhance coverage, illumination, and received signal strength. Figure~\ref{adt} (b) illustrates how the LC-based RIS can provide dynamic beam steering, which is useful in indoor environments where users are mobile. Figure~\ref{adt} (c) demonstrates the joint light amplification and beam focusing capabilities of an LC-based RIS-enabled VLC transmitter. Finally, Fig.~\ref{adt} (d) highlights how the LC-based RIS-enabled VLC transmitter can produce light beams with narrow beamwidth to reduce interference, provide multiple beams with different intensities via controlled amplification and beam focusing to support differentiated quality-of-service requirements for multiple users, and enhance the security of indoor VLC systems. These application scenarios reveal how higher communication (e.g., energy efficiency, data rate, and security) and illumination performances can be achieved without any additional resources, i.e., no extra transmit power or bandwidth resources. Moreover, this novel RIS-enabled VLC transmitter can assist in addressing the LoS blockage issue through light amplification and beam focusing onto an RIS (e.g., mirror array) in the channel.
\vspace*{-3mm}

 \subsection{Channel Model, Data Rate, and Illumination Expression}
The channel model of a VLC system with an LC-based
RIS-enabled transmitter characterizes the propagation of light
beams from the LED arrays, through the LC-based RIS in
front of the LED arrays, and finally through air to the PD
of the VLC receiver. The channel gain can be expressed as
$H=\alpha_{LC} \times {G_{\rm LoS}}$, where $\alpha_{LC}$ denotes the LC’s transmission coefficient defined in \cite{9910023} and ${G_{\rm LoS}}$ is the direct current gain of the LoS propagation \cite{9799770}. The LC's transmission coefficient can be tuned by optimizing the refractive index of the LC in the presence of an external electric field to control the emerged light direction as revealed in \cite{9910023}. Note that for the VLC transmitters (i.e., LED array arrangements) in Fig.~\ref{figs01}, $H={G_{\rm LoS}}$. 

The achievable data rate for any user in an indoor environment served by a VLC system with an LC-based RIS-enabled
transmitter with the channel gain, $H$, the optical transmit power,
$P$, and the amplification gain coefficient, $\Gamma$, which is defined in \cite{R14} can be determined using the rate expression in \cite{9910023}.

Uniform lighting distribution is essential to the reliability of the VLC systems and, as a result, needs to be considered in RIS-aided VLC systems. The reason is to ensure that the RIS placed in front of the LED arrays does not degrade the illumination properties of the VLC transmitter or cause any health risk. The illumination intensity at a surface inside a room with an LC-based RIS-enabled transmitter can be measured by the surface illuminance which is expressed as \cite{9799770,9910023}\\

$I=\exp\left({\Gamma d}\right) \times P \times \alpha_{\rm LC} \times \frac{{\left( {m + 1} \right)}}{{2\pi {l^2}\delta }}{\cos ^m}\left( {{\Phi }} \right)\cos \left( {{\varphi}} \right)G\left( {{\varphi}} \right),$\\\\
where $\Gamma$ denotes the amplification gain coefficient defined in \cite{9910023}, $\alpha_{\rm LC}$ is the transition coefficient, $m =  - {{{{\log }_2}\big( {\cos \big( {{\phi_{{1}/{2}}}} \big)} \big)}}^{-1}$ is  the Lambertian emission order with ${{\phi_{{1}/{2}}}}$ as the LED's semi--angle at half power, $l$ is the distance between the transmitter and the surface, $\delta$ is the optical to luminous flux conversion factor, ${\Phi}$ is the angle of irradiance, $\varphi$ is the angle of incidence, and $G\left( {{\varphi}} \right)=f^2/\sin^2\varphi, 0\le\varphi\le{\phi_{{1}/{2}}}$ is the gain of the non-imaging concentrator which focuses the light from the LEDs into the LC-based RIS with ${\phi_{{1}/{2}}}$ as the LED's semi-angle at half power and $f$ as the internal refractive index of the concentrator. In this expression, $P$ is the optical transmit power from the LED array which is incident on the LC-based RIS and $\exp\left({\Gamma d}\right) \times P \times \alpha_{\rm LC}$  represents the emerged optical power from the LC-based RIS after the incident light has propagated through the LC module and undergone amplification in the presence of an external electric field. The remaining part of the expression denotes the luminous flux of a unit optical power.
\vspace*{-3mm}
\begin{figure*}[!t]
    \centering
    {{\includegraphics[width=0.75\textwidth]{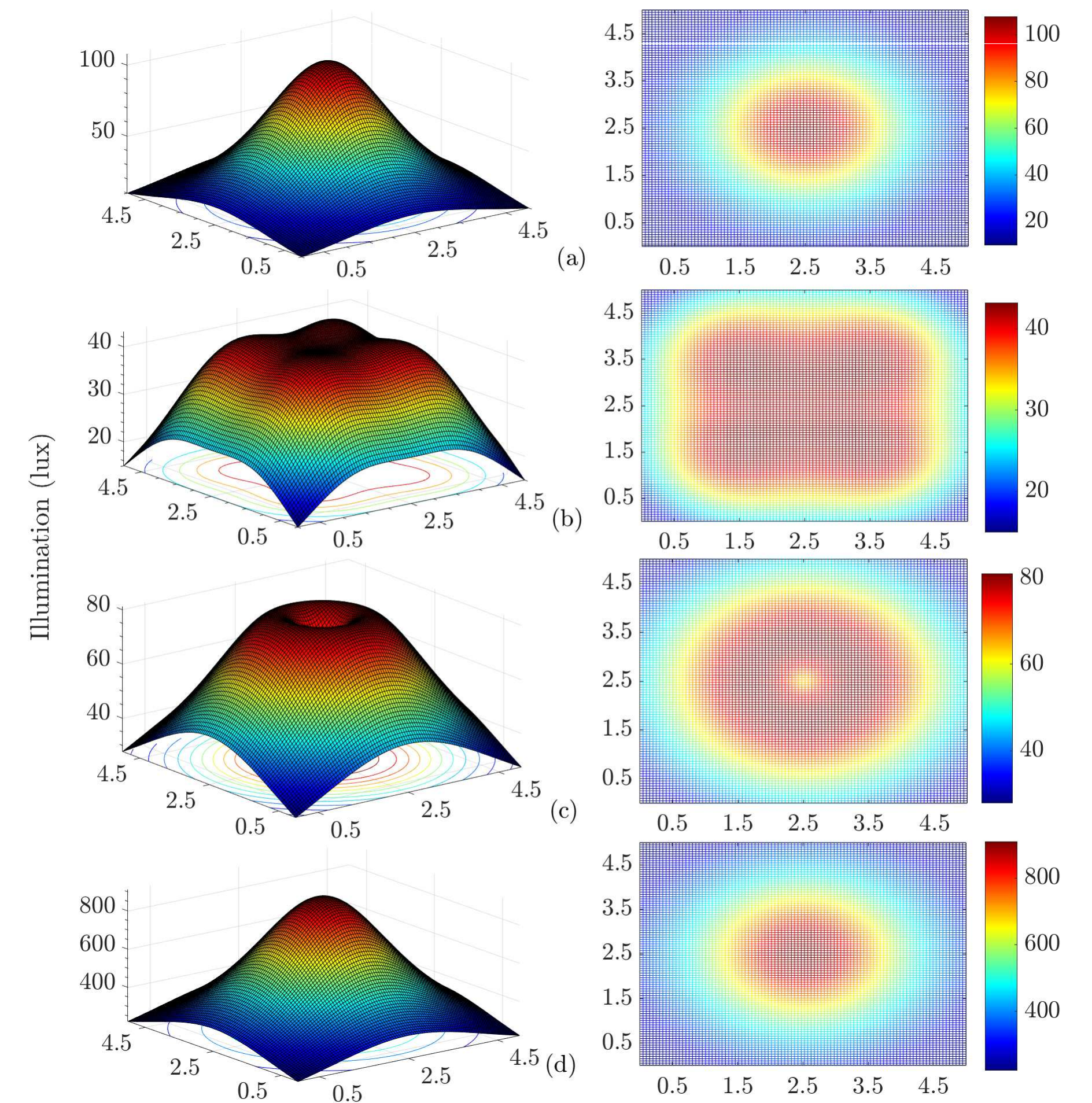}}}
    \caption{\small{Illumination distribution comparison for VLC transmitters characterized by different LED array arrangement and LC-based RIS-enabled VLC transmitter: (a) centralized LED array arrangement (Min. 10 lux, Max. 107 lux); (b) distributed LED array arrangement (Min. 15 lux, Max. 43 lux); (c) ADT LED array placement (Min. 28 lux, Max. 81 lux); (d) proposed centralized LED array with LC-based RIS (Min. 222 lux, Max. 910 lux). {\color{black}{The $x$, $y$, and $z$ axes represent the length (m), width (m), and illumination (lux), respectively.}}}}%
  \label{figsp2}
    \vspace*{-4mm}
\end{figure*}

\section{Analysis of Achievable Data Rate and Illumination Performance}
In this section, a performance comparison of the different LED arrays arrangements and the proposed RIS-enabled VLC transmitter is performed. The performance metrics used in our analysis are the achievable data rate, surface illumination, illumination uniformity, and data rate uniformity. Illumination uniformity can be defined as the ratio between the minimum and the average illumination among all surfaces or sensing points in an indoor environment \cite{9799770,9037594,luxst}. Similarly, data rate uniformity refers to the ratio between the minimum and the average data rate among all surfaces or users. According to \cite{9037594},  values close to 1 indicate uniform lighting or data rate condition and, in general, values more than 0.7 are desired. However, a uniformity value of 0.4 is considered the least acceptable for illumination purposes \cite{luxst}.  

Without loss of generality, a $5$ m $\times$ $5$ m $\times$ $3$ m room size is considered \cite{7096280,Wang12,9799770,9037594}. For the distributed LED array placement scheme, the entire room is assumed to be divided into four equal quadrants, each with an LED array ceiling lamp at the center. For the remaining LED array arrangement schemes and the proposed approach, the LED arrays are deployed at the center of the room. The illumination and data rates are measured at a typical desktop height (i.e., receiver plane) of 0.85 m above the floor. To better determine the performances of the various schemes across the whole room, the receiver plane is divided into  
$100\,\times 100\,$ grid points which serve as sensing points. This is in accordance with the European lighting standard \cite{luxst}, which requires the use of the grid specification to calculate and measure illumination averages and uniformity. For the considered simulations, the information carrying bandwidth is set as 200 MHz, $\tau = 45^{\circ}$, $d=0.75$ mm, $v_{\rm th}=1.34$ V, and $P=1$ W. The LEDs operate at a wavelength of 510 nm. The analysis for the RIS-enabled transmitter is carried out for an LC-based RIS element with a fixed refractive index $n_c=1.55$ ($v_{\rm e}=2.1$ V is the required voltage). All other simulation parameters are set according to \cite{9910023}. 

\begin{figure*}[!t]
    \centering
    {{\includegraphics[width=0.75\textwidth]{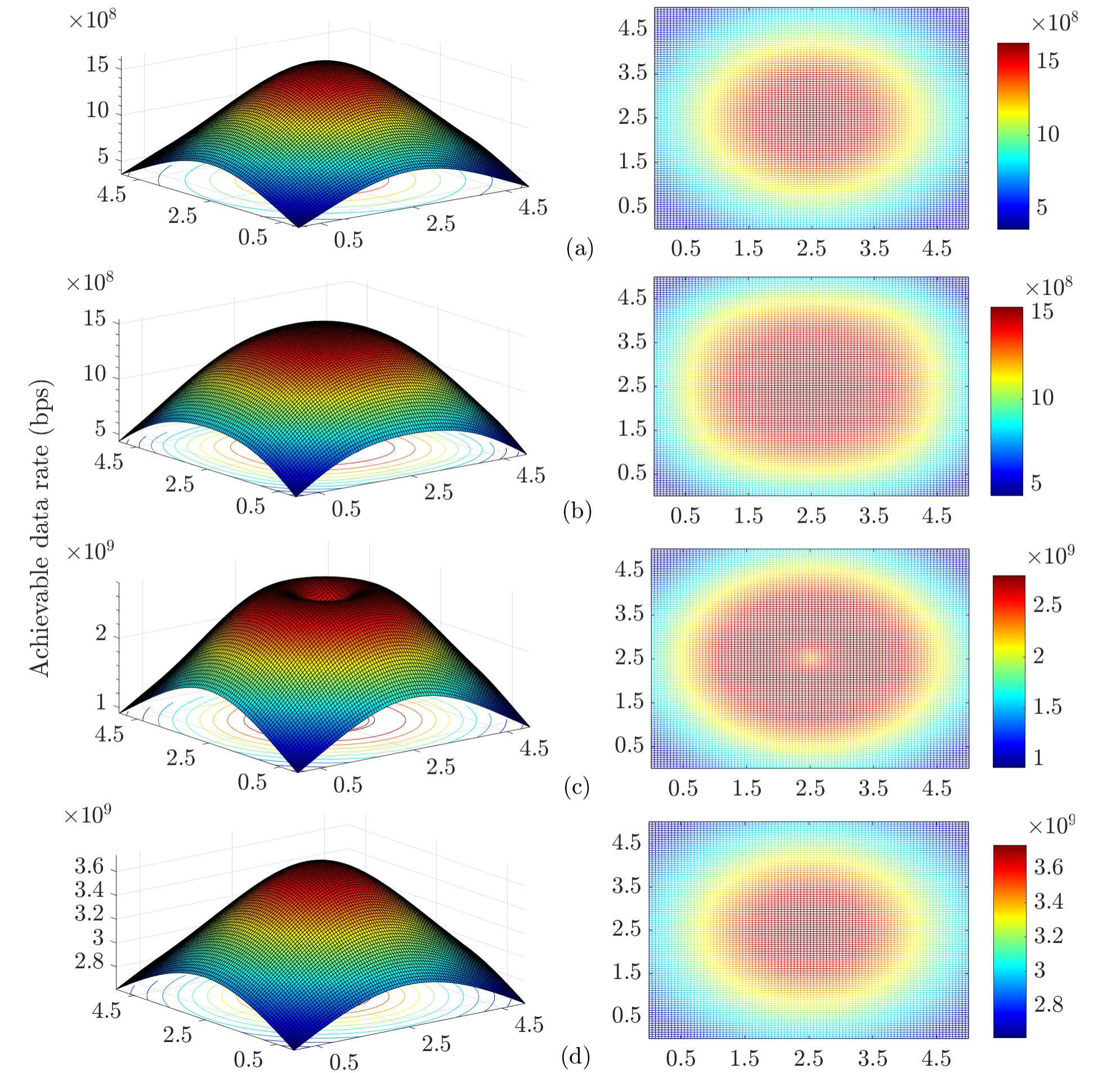}}}
    \caption{\small{Data rate distribution comparison for VLC transmitters characterized by different LED array arrangement and an LC-based RIS-enabled VLC transmitter: (a) centralized LED array arrangement (Min. 0.3595 Gbps, Max. 1.6399 Gbps); (b) distributed LED array arrangement (Min. 0.4351 Gbps, Max. 1.5373 Gbps); (c) ADT LED array placement (Min. 0.9179 Gbps, Max. 2.8024 Gbps); (d) proposed centralized LED array with LC-based RIS (Min. 2.6115 Gbps, Max. 3.7363 Gbps). {\color{black}{The $x$, $y$, and $z$ axes represent the length (m), width (m), and data rate (bps), respectively.}}}}%
  \label{figsp3}
    \vspace*{-4mm}
\end{figure*}
Figure~\ref{figsp2} compares the illumination distribution in an indoor environment for a VLC system with a transmitter characterized by typical LED array arrangement schemes with that of the proposed RIS-enabled VLC transmitter. This figure is generated by partitioning the receiver plane into small grids and determining the illumination in each grid. Several key observations can be made from Fig.~\ref{figsp2}. To begin with, the illumination is highest at the center of the room and lowest at the corners for all the different LED array arrangement schemes and the proposed transmitter. This observation can be explained by the Lambertian radiation pattern of LEDs. The highest illumination for the centralized, distributed, and ADT arrangement schemes are significantly below the recommended minimum indoor lighting requirement of 150 lux for any office/gym/home activities as described in \cite{luxst}. As can be observed, the illumination at the task or activity area, which is typically the center of the room, and the room corners is very low. As a result, these approaches may not always satisfy the illumination function in VLC. However, this is not the case for the proposed transmitter, as illumination is greatly enhanced due to the light amplification and beam focusing capabilities of the LC-based RIS. The minimum illumination requirement is satisfied for the task or activity area and the corners of the room. Secondly, the proposed approach satisfies the minimum required illumination of 400 lux for reading/office work according to the European lighting standard \cite{luxst}, as more than 50$\%$ of the total room area has illumination above the threshold of 400 lux. Moreover, the illumination of the immediate surrounding area of the task area and that of the background area fall within the recommended values (i.e., over $500$ lux for the immediate surrounding and above 200 lux for the background area).\footnote{According to the European lighting standard \cite{luxst}, the immediate surrounding area refers to the space just around the task or activity area and the area outside the immediate surrounding is referred to as the background area. The illumination of the immediate surrounding area should be related to the illumination of the task area, and the illumination of the background area should be related to that of the immediate surrounding to avoid significant spatial variations in indoor illumination.} Thus, it is acceptable that the illumination of the room corners is lower than that of the task or activity area. The above analyses reveal that the proposed LC-based RIS-enabled transmitter has superior illumination performance and fulfills the visual comfort needs. Furthermore, compared with the centralized LED array arrangement scheme (i.e., without the LC-based RIS), up to 2100$\%$ gain in illumination is obtained with the deployment of an RIS at the transmitter side. Note that a similar performance gain is expected when an LC-based RIS is used with the distributed and ADT LED array arrangement schemes. In addition, the proposed approach obtains illumination performance gain of up to 2000$\%$ and  1325$\%$ when compared to the distributed and ADT placement schemes, respectively.

Figure~\ref{figsp3} illustrates the data rate distribution for the considered LED array arrangement schemes and the proposed RIS-enabled VLC transmitter. {{This figure is generated by partitioning the receiver plane into small grids and determining the data rate in each grid.}} It can be observed from this figure that the proposed approach achieves up to 626$\%$, 500$\%$, and 184$\%$ improvement in data rate when compared to the centralized, distributed, and ADT LED array arrangement schemes, respectively. Moreover, only the proposed approach has a minimum data rate on the order of Gbps, as the remaining approaches could only support Mbps download speeds. This reveals that all users can enjoy Gbps download speed irrespective of their location (i.e., in the corner of the room or at the center) for the same number of LED arrays, transmit power, and bandwidth resources. Thus, in addition to the significant increase in the peak data rates, higher data rates can be guaranteed for the vast majority of the locations in the indoor environment. This significant performance enhancement is due to the new reconfigurable and transmit light amplification capability of the RIS-enabled transmitter.  

Table~\ref{tab1} shows the illumination and data rate uniformity  for the different LED array arrangement schemes and the proposed approach based on the results in Figs.~\ref{figsp2} and \ref{figsp3}. {{The data rate (illumination) uniformity value for any LED array arrangement scheme is the ratio between the minimum value of the data rates (illumination) for all the grid points and the average data rate (average illumination) among all the grid points.}} The centralized LED array arrangement has the worst illumination and data rate uniformity. Its uniformity values of 0.2371 and 0.3535 are below the acceptable values of 0.40 for illumination and 0.7 for data rate, respectively. Hence, the centralized scheme is characterized by high illumination and data rate fluctuations over the entire room and exhibits an unacceptable performance level in achieving the dual role of communication and illumination. The two remaining LED array arrangement schemes and the proposed approach demonstrate uniform illumination and, hence, would experience the least fluctuation. Besides, these uniformity values (i.e., 0.4378, 0.4755, and 0.4628) are practical since it is acceptable for some parts of the room space to have lower illumination. Moreover, a comparison of the data rate uniformity values of 0.3937, 0.4379, and 0.8168 for the distributed, ADT, and the proposed schemes, respectively, suggests that the proposed approach can support high data rates across the entire room. More reliable and high quality communication can be guaranteed for all users in an indoor environment with the proposed RIS-enabled transmitter. Unlike illumination,  data rate uniformity values of 0.3535, 0.3937, and 0.4379 are not acceptable for VLC. For such values, the data demands of all users may not be satisfied since the achievable data rate becomes highly dependent on the user's location (i.e., extremely high data rate for users in the center of the room and extremely low data rate for users far away from the center).

\begin{table}[]
\centering
\caption{Illumination and data rate uniformity for indoors.}
\label{tab1}
\begin{tabular}{|l||ll|}
\hline
\multirow{2}{*}{\textbf{LED Array Arrangement Type}} & \multicolumn{2}{l|}{\textbf{Uniformity Distribution}} \\ \cline{2-3} 
                       & \multicolumn{1}{l|}{\textbf{Illumination}} & {\textbf{Data Rate}} \\ \hline\hline
Centralized LED Array & \multicolumn{1}{l|}{0.2371}       & 0.3535    \\ \hline
Distributed LED Array & \multicolumn{1}{l|}{0.4378}       & 0.3937    \\ \hline
ADT LED Array         & \multicolumn{1}{l|}{0.4755}       & 0.4379    \\ \hline
Proposed               & \multicolumn{1}{l|}{0.4628}       & 0.8168    \\ \hline
\end{tabular}
  \vspace*{-3mm}
\end{table}

\begin{figure}
    \centering
    {{\includegraphics[width=0.5\textwidth]{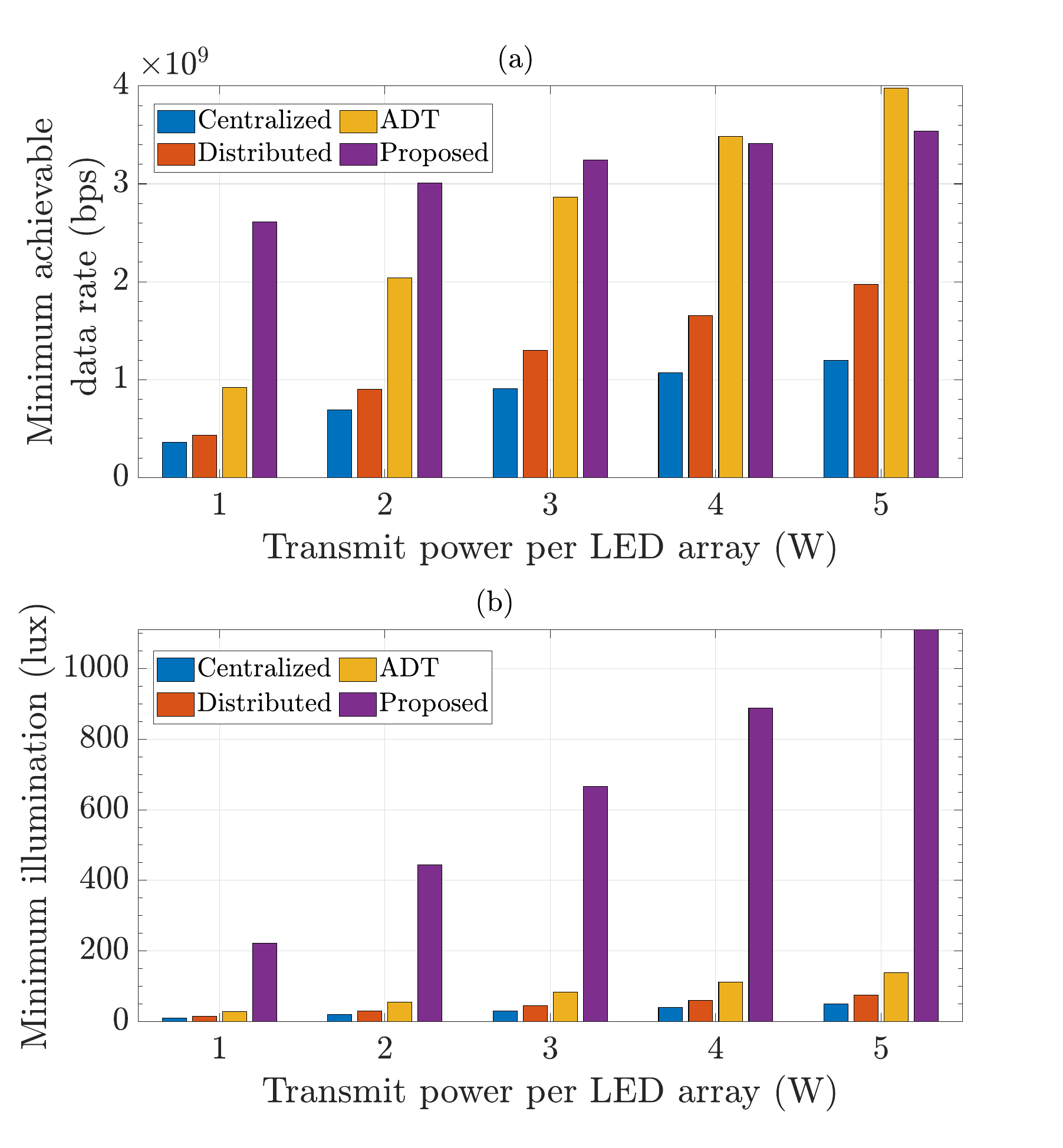}}}
    \caption{\small{Impact of transmit power on the performance of a VLC system for the considered approaches: (a) minimum achievable data rate vs. transmit power; (b) minimum illumination vs. transmit power.}}%
  \label{figsa2}
  \vspace*{-3mm}
\end{figure}

Finally, Fig.~\ref{figsa2} illustrates the achievable data rate and illumination performance of an indoor VLC system for different transmit power values. In this figure, the minimum data rate and illumination have been plotted for different transmit power values for the centralized, distributed, ADT, and the proposed LED array arrangement schemes. It can be observed that the minimum data rate and minimum illumination increase with the transmit power for all the considered approaches. However, growth trends for the data rate and illumination distribution are different because, unlike surface illumination, data rate is a log function. The proposed design outperforms the remaining approaches in terms of mimimum data rate when the transmit power is less than 4 W. For transmit power values of   4 W and above, the ADT LED array arrangement scheme achieves a higher minimum data rate than the LC-based RIS-enabled transmitter. The reason for this observation is that, while the data rate for the proposed scheme begins to saturate, that of the ADT approach keeps increasing for transmit power values greater than 4 W. This growth trend of the ADT approach, which is similar to that of the distributed LED array arrangement, is due to the spatial configuration of its LED arrays and the fact that the received power depends on the position of the LED arrays and the angle of irradiance.  Although the ADT LED arrays arrangement scheme demonstrates the best minimum data rate performance for transmit power values above 4 W, its illumination remains worse than that of the proposed approach. Thus, even for higher transmit power values, the centralized, distributed, and ADT LED array arrangement schemes fail to provide acceptable joint data rate and illumination performance.
Note that the proposed approach involves placing the LC-based RIS in front of the centralized LED array arrangement scheme. It can be inferred from the simulation results that higher performance gains could be expected when the LC-based RIS is placed in front of the ADT LED arrays since the ADT arrangement scheme outperforms the centralized LED arrays placement scheme. 

\section{Open Research Opportunities and Challenges}
Unlike the deployment of RISs in the channel, which has received great research attention, the application of RISs in VLC transceiver design and performance optimization has yet to be explored. This article has demonstrated the great potential of LC-based RIS-enabled VLC transmitter. By presenting a novel approach of using LC-based RIS to (i) achieve a much higher level of illumination, (ii) significantly improve data rate, and (iii) jointly enhance illumination and data rate uniformity, the following promising research opportunities and challenges have been identified:

{\textbf{1. Performance Optimization under Joint Illumination and Communication Constraints:}} The performance of LC-based RISs is highly dependent on the electronic tuning of the refractive index since its value determines both the amplification and transmission gain coefficients of the LC-based RIS. Thus, the expression for the achievable data rate in VLC systems with an  RIS-enabled transmitter differs from a VLC system without the RIS. Moreover, it is vital to consider the joint optimization of the LC's refractive index and the available transmit power of the VLC system to enhance the illumination and communication performances further. Several objective functions such as data rate maximization, energy efficiency maximization, load balancing, max-min rate, transmit power minimization, and secrecy rate maximization need to be examined under illumination and communication related constraints (e.g., quality-of-service, quality-of-experience, and LC-based RIS tuning time). Furthermore, novel constraints related to the amplification gain should be studied as a network operator or a customer might want to emphasize or de-emphasize the amplification capability of the LC-based RIS-enabled VLC transmitter. Due to the resulting unique channel gain, illumination, and data rate expressions, the potential high dimensionality of the optimization problems, and the new design constraints, traditional approaches for optimizing the performance of VLC systems cannot be directly adopted. New mathematical optimization techniques with low complexity and machine learning-driven approaches should be explored in further research, proof-of-concept studies, and, finally, the realization of LC-based RIS-enabled transmitters in VLC. These developments could be extended into other VLC applications, such as vehicle-to-vehicle and underwater wireless communication since their channel models differ from that of the indoor environment.

{\bf{2. Analysis of the Impact of VLC and RIS Parameters on Illumination and Communication Performance:}} VLC transmitters and systems generally have several design parameters such as the LED's semi-angle at half power, available transmit power and subchannels, room dimension and shape, wall reflection coefficients, and the distance between LED arrays. On the other hand, an LC-based RIS element has parameters that can affect its performance gains. Such parameters include the threshold voltage, thickness of the LC, and transmission wavelength. Due to page limitations, this article considered fixed values for the parameters mentioned above. It would be worth examining how different system parameters will impact the system performance and potentially yield further improvements in data rate and illumination. Moreover, optimization problems with such parameters as decision variables need to be considered to determine the optimal values for various performance metrics. 

{\bf{3. Noise Effect on the Performance of the LC-based RIS:}} The LC-based RIS performs transmit light amplification through the process of stimulated emission that generates new photons. In practice, the photon generation process can limit the data rate and illumination performances. This is because tuning LCs takes time and is characterized by the response time (which is typically in the order of milliseconds). In addition, the photon generation process might result in noise affecting the system's performance. It is therefore important to investigate ways of decreasing the response time, how often to perform LC tuning, and noise effects on the performance of LC-based RIS-enabled VLC transmitters.

{\bf{4. Adaptive Cell Formation for VLC:}} The ability of LC-based RIS-enabled transmitters to focus, amplify, and steer emitted light due to their electronically adjustable refractive index make them suitable for the design of adaptive cell formation algorithms. Unlike ADTs, where the resulting cells formed by the highly directional LED arrays cannot be altered with changes in the indoor environment (e.g., change in user density and distribution, data rate and illumination requirements, and mobility), LC-based RIS-enabled transmitters permit the design of algorithms to control both the refractive index and transmit power to move illumination coverage areas in indoors and create adaptive communication cells. The impact of such joint power allocation and cell formation algorithms on the illumination and communication performance of LC-based RIS-enabled VLC systems is another exciting research area for future work.

{\bf{5. Impact on Cost of VLC:}} The development of RISs, in general, and optical RISs, in particular, is still at an early stage. As a result, much has yet to be known about the cost of RISs. Inserting an LC-based RIS into an LED package can affect its cost. Since low deployment cost is one of the key features that network operators look out for, it is critical to investigate how RISs would affect the cost of LEDs. Such investigations should focus on theoretical studies and practical deployments that consider imperfections in the system model to accurately examine the cost-benefit analysis. 

{\color{black}{{\bf{6. Enhanced VLC for joint communication, illumination, and sensing:}} VLC has been identified as a key technology for indoor joint communication and sensing. The quality of the received signal plays a critical role in the communication performance and sensing accuracy of such an integrated sensing and communication system. The signal focusing and amplification capabilities of the  proposed  LC-based RIS-enabled VLC transmitter can be leveraged to enhance the received signal strength in VLC-based integrated sensing and communication systems.}} 


\section{Conclusion}
This article investigated a novel approach of using LC-based RIS to improve the data rate and illumination distribution in indoor VLC systems. Specifically, the illumination and data rate performances of popular approaches for LED array arrangement were first analyzed. Then, a novel VLC transmitter design that involves placing LC-based RIS in front of the LED arrays was introduced, and the channel gain and data rate expressions were presented. Simulation results revealed that, unlike popular LED array arrangement schemes, the proposed LC-based RIS-enabled VLC transmitter achieves a higher data rate and illumination distribution. Moreover, it enhances data rate uniformity and demonstrates that a much higher level of data rate and illumination can be jointly obtained from the same number of LED arrays and transmit power value when an LC-based RIS with a 1.34 V - 10 mA source is employed. Thus, more energy efficient RIS-aided VLC systems can be designed due to the nearly-passive nature of the LC-based RIS. Furthermore, several open challenges have been highlighted. Since the use of RISs at the transmitter side in VLC remains unexplored, this article can provide a helpful guide and inspire further studies on optical RIS-aided communications.







\bibliographystyle{IEEEtran}
\bibliography{IEEEfull}

\end{document}